\documentclass[aip,apl,reprint]{revtex4-1}

\usepackage[T1]{fontenc}
\usepackage{lmodern}
\usepackage{graphicx}
\usepackage{dcolumn}
\usepackage{bm}

\usepackage[utf8]{inputenc}
\usepackage[T1]{fontenc}
\usepackage{mathptmx}
\usepackage{etoolbox}
\usepackage{upgreek} 
\usepackage{multirow} 
\usepackage{newtxmath} 

\usepackage{bbm}
\usepackage{amsmath}
\usepackage[english]{babel}
\usepackage{ragged2e}
\usepackage[separate-uncertainty=true]{siunitx}
\usepackage{hyperref}
\hypersetup{
    colorlinks=true,allcolors=blue,}
\graphicspath{{Figs/}}

\makeatletter
\def\@email#1#2{%
 \endgroup
 \patchcmd{\titleblock@produce}
  {\frontmatter@RRAPformat}
  {\frontmatter@RRAPformat{\produce@RRAP{*#1\href{mailto:#2}{#2}}}\frontmatter@RRAPformat}
  {}{}
}%
\makeatother
\begin{document}

\preprint{AIP/123-QED}

\date{\today}

\title{Vector Magnetometry Using Shallow Implanted NV Centers in Diamond with Waveguide-Assisted Dipole Excitation and Readout}
\author{Sajedeh Shahbazi}
\affiliation{Institute for Quantum Optics, Ulm University, D-89081 Ulm, Germany}

\author{Giulio Coccia}
\affiliation{Institute for Photonics and Nanotechnologies (IFN) - CNR, Piazza Leonardo da Vinci, 32, Milano 20133, Italy}

\author{Argyro N. Giakoumaki}
\affiliation{Institute for Photonics and Nanotechnologies (IFN) - CNR, Piazza Leonardo da Vinci, 32, Milano 20133, Italy}

\author{Johannes Lang}
\affiliation{Institute for Quantum Optics, Ulm University, D-89081 Ulm, Germany}

\author{Vibhav Bharadwaj}
\affiliation{Institute for Quantum Optics, Ulm University, D-89081 Ulm, Germany}
\affiliation{Department of Physics, Indian Institute of Technology Guwahati, 781039 Guwahati, Assam, India}

\author{Fedor Jelezko}
\affiliation{Institute for Quantum Optics, Ulm University, D-89081 Ulm, Germany}
\affiliation{Center for Integrated Quantum Science and Technology (IQst), Ulm University, Ulm D-89081, Germany}

\author{Roberta Ramponi}
\affiliation{Institute for Photonics and Nanotechnologies (IFN) - CNR, Piazza Leonardo da Vinci, 32, Milano 20133, Italy}

\author{Anthony J. Bennett}
\affiliation{School of Engineering, Cardiff University, Queen's Building, The Parade, Cardiff, CF24 3AA, UK}
\affiliation{Translational Research Hub, Cardiff University, Maindy Road, Cardiff, CF24 4HQ, United Kingdom}

\author{John P. Hadden}
\affiliation{School of Engineering, Cardiff University, Queen's Building, The Parade, Cardiff, CF24 3AA, UK}
\affiliation{Translational Research Hub, Cardiff University, Maindy Road, Cardiff, CF24 4HQ, United Kingdom}

\author{Shane M. Eaton}
\affiliation{Institute for Photonics and Nanotechnologies (IFN) - CNR, Piazza Leonardo da Vinci, 32, Milano 20133, Italy}

\author{Alexander Kubanek}
\altaffiliation[Corresponding author: ]{alexander.kubanek@uni-ulm.de}
\affiliation{Institute for Quantum Optics, Ulm University, D-89081 Ulm, Germany}
\affiliation{Center for Integrated Quantum Science and Technology (IQst), Ulm University, Ulm D-89081, Germany}

\date{\today}

\begin{abstract}

On-chip magnetic field sensing with Nitrogen-Vacancy (NV) centers in diamond requires scalable integration of 3D waveguides into diamond substrates. Here, we develop a sensing array device with an ensemble of shallow implanted NV centers integrated with arrays of laser-written waveguides for excitation and readout of NV signals. Our approach enables an easy-to-operate on-chip magnetometer with a pixel size proportional to the Gaussian mode area of each waveguide. The performed continuous wave optically detected magnetic resonance on each waveguide gives an average dc-sensitivity value of $\SI{195 \pm 3}{\mathrm{nT}/\sqrt{\mathrm{Hz}}}$, which can be improved with lock-in-detection or pulsed-microwave sequences. We apply a magnetic field to separate the four NV crystallographic orientations of the magnetic resonance and then utilize a DC current through a straight wire antenna close to the waveguide to prove the sensor capabilities of our device. We reconstruct the complete vector magnetic field in the NV crystal frame using three different NV crystallographic orientations. By knowing the polarization axis of the waveguide mode, we project the magnetic field vector into the lab frame.   

\end{abstract}

\maketitle

\section{\label{sec:Intro}Introduction}

Negatively charged Nitrogen-Vacancy ($\mathrm{NV}$) centers in diamond have attracted significant attention for their remarkable potential in quantum sensing, imaging, and quantum information processing~\cite{10.1063/1.2943282,steinert_high_2010, davis_mapping_2018, doherty_nitrogen-vacancy_2013}. In quantum sensing, ensembles of NV centers are a promising platform for continuously sensing static and low-frequency magnetic fields \cite{hong_nanoscale_2013} and their chemical inertness enables direct physical contact with delicate biological systems~\cite{schirhagl_nitrogen-vacancy_2014}. Additionally, shallow-implanted NV centers, created via ion implantation~\cite{ofori-okai_spin_2012} or thin film growth~\cite{ohashi_negatively_2013}, enhance sensitivity by placing them in close proximity to samples~\cite{janitz_diamond_2022}. Another advantage of using NV ensembles is the ability to measure the projection of the magnetic field along each of the four distinct NV symmetry axes to form a 3D vector field \cite{maertz_vector_2010}.  

Integrating photonics to address large ensembles of NV centers presents significant advantages, optimizing the scalability and efficiency of these quantum systems~\cite{elshaari_hybrid_2020, johnson_diamond_2017}. A promising integrated platform is laser written waveguides in diamond~\cite{bharadwaj_femtosecond_2019, sotillo_diamond_2016,eaton_quantum_2019}. These waveguides are generated using tightly focused femtosecond laser pulses, creating polarized single-mode waveguide intersecting with the shallow-implanted NV centers. Our integration approach combines shallow-implanted NV ensemble with laser-written waveguides inside diamond, forming a two-dimensional sensing array. This technique enables control over the light propagation path, facilitating excitation and fluorescence detection from a large ensemble of NV centers~\cite{hoese_integrated_2021}.  The excitation and fluorescence collection of an ensemble of shallow-implanted NV centers via laser-written waveguides offers a pathway toward scalable, efficient, and easy-to-use quantum sensing devices.
\\
The paper is structured as follows. In Section~\ref{characterization}, we characterize the shallow implanted NV centers using a home-built confocal microscope. We calculate the magnetic field sensitivity and coherence properties of the NVs by conducting continuous wave optically detected magnetic resonance (CW-ODMR) and pulsed measurements. We then characterize the sensitivity performance using waveguide-assisted ODMR. In Section~\ref{sec-vector-sensing}, to demonstrate the sensing capabilities of our NV centers via the waveguides, we apply an external DC current and measure the resonance shift in the ODMR spectrum. For reconstruction of the magnetic vector field, we use three different NV crystallographic orientations and, by knowing the waveguide's polarization axis alongside the polarization properties of the different NV families, we derive the full vector field in both the crystal frame and the lab frame. We describe the experimental setup in Section~\ref{sec-setup} and conclude in Section~\ref{sec-summary}.

\section{experiment}\label{sec-experiment}
\subsection{NV ensemble characterization}\label{characterization}

The diamond is a type-IIa electronic-grade slab ($\SI{2}{mm} \times \SI{2}{mm} \times \SI{0.5}{mm}$) with the surface cut along (100) and the side facets along (110). The type-II waveguides are created using femtosecond pulsed laser at $\lambda = \SI{515}{nm}$ with a repetition rate of $\SI{500}{kHz}$. The laser beam is focused into the sample with a high-numerical-aperture (NA) objective (1.25 NA, $100\times$), creating waveguide sidewalls with cross section of approximately $\SI{5}{\mu m}$ transveresley and $\SI{20}{\mu m}$ vertically with $\SI{2}{mm}$ length according to the size of the diamond. Using three different powers of laser at 30, 40, and 50 mW, creates waveguide arrays at three different depths (40, 75 and $\SI{110}{\mu m}$) from the surface. The waveguide walls are separated by $\SI{13}{\mu m}$ optimized for single mode guiding of NV spectrum emission from 630 to $\SI{800}{nm}$~\cite{sotillo_diamond_2016}. This waveguide geometry results in a Gaussian mode when coupling to the waveguide~\cite{hoese_integrated_2021}. Afterward, we shallow implant the NV ensemble on the diamond's side facet (110) with an implantation depth 8-12 nm below the surface and achieving an approximate NV density of $75 \pm 25 \times 10^{9}$  $\text{NV}/\mathrm{cm^2}$~\cite{pezzagna_creation_2010}. (See Section~\ref{sec-setup} for additional information regarding the fabrication method and experimental setup.)

\begin{figure}
\includegraphics[scale=0.68]{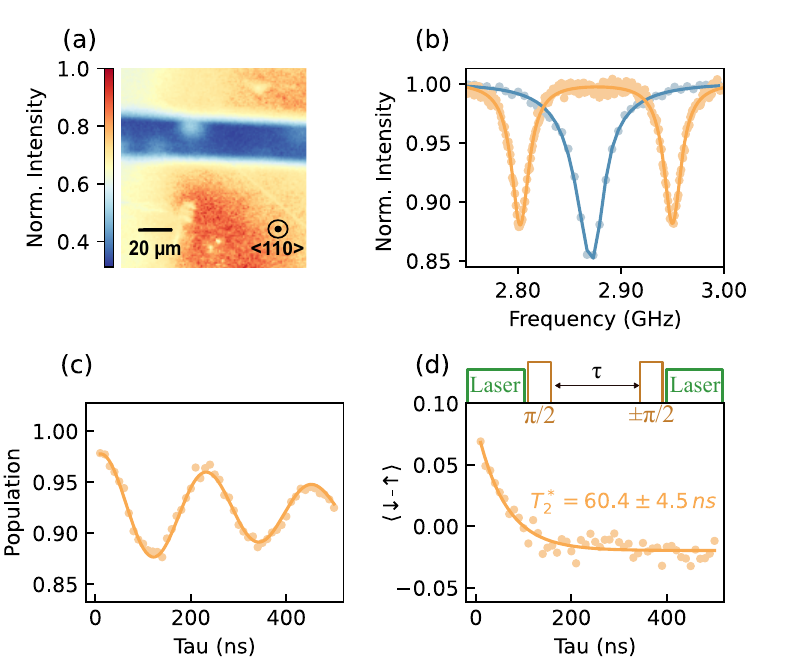} 
\caption{ \justifying \textbf{Testing NV-based sensing via confocal free-space access.} (a) Confocal image of the shallow implanted NV ensemble. The shaded area is a straight copper wire that applies the MW field to the NVs. (b) CW-ODMR in confocal with (orange) and without (blue) an external B-field along \(\langle 100 \rangle\) direction. The scattered data points are fitted with a single (blue) and double (orange) Lorenzian function. (c) Rabi oscillation between ${m_s=0}$ and ${m_s=+1}$ at $\SI{2.95}{GHz}$ microwave resonance frequency. The Rabi oscillation has a $\SI{4.54 \pm 0.03}{MHz}$ frequency (from fitting) and a lifetime of $\SI{570 \pm 50}{ns}$. (d) The Ramsey free induction decay with \SI{111}{ns} $\pi/2$ time gives a dephasing time of $T_2^*=\SI{60.4 \pm 4.5}{ns}$ corresponding to optimal sensing time. The schematic of the pulse sequence can be seen on top of the graph.} \label{fig-one}
\end{figure}

\begin{figure*}
\centering
\includegraphics[scale=0.75]{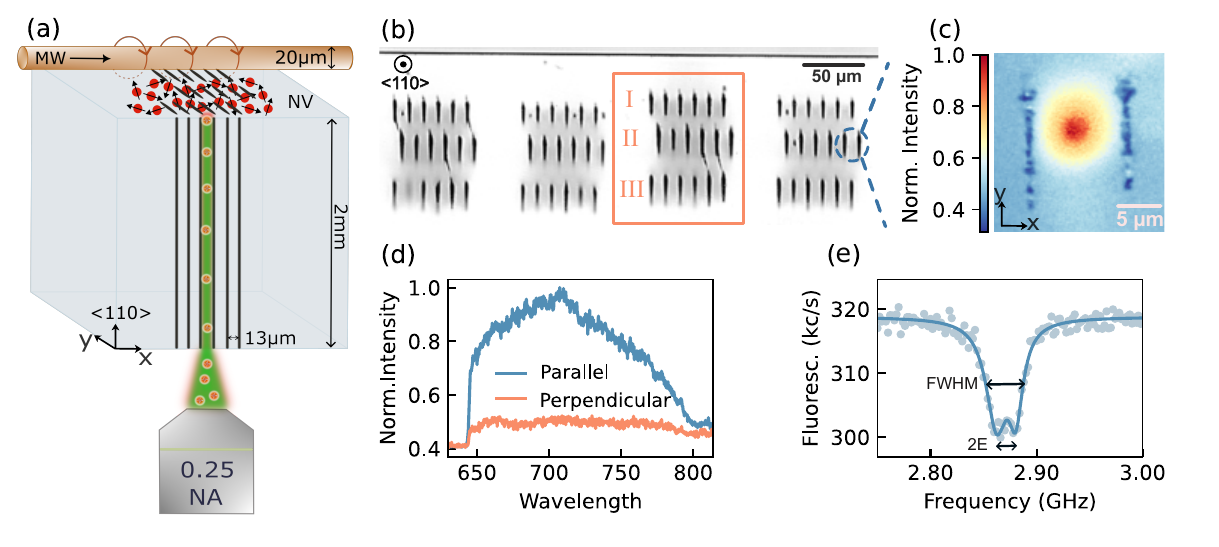} 
\caption{\justifying \textbf{Characterization of the laser written waveguide arrays and their sensing performance.}(a) A schematic of the waveguide-assisted sensor array with shallow-implanted NV centers is presented. The optical excitation and fluorescence detection is mediated via the laser written waveguides and addressed using a low-numerical-aperture (low-NA) objective. (b) Microscope image of a 2D waveguide array looking at the surface where the NV centers are implanted. For the waveguide set marked with an orange box, the dc-sensitivity values derived from Eq.~\eqref{eq:sensitivity} are shown in Table~\ref{your-table-i} in the main text. (c) Confocal image of a single waveguide with a laser coupled to the waveguide mode depicting a Gaussian mode with $1/e^2$ area of 456 ${\mathrm{\mu m^2}}$. The $y$-axis is along the polarization axis of the waveguide. (d) The NV PL spectrum through excitation and detection via a waveguide with light's polarization parallel (blue) and perpendicular (orange) to the polarization axis of the waveguide mode. (e) Waveguide-assisted CW-ODMR from one of the studied waveguides.}\label{fig-two}
\end{figure*}

We first characterize the NV ensemble sensitivity performance and coherence properties in a confocal microscope using CW-ODMR and pulsed sequences. CW-ODMR is a simple and powerful technique in which a continuous microwave is swept over the spin resonance of one or multiple NV centers while the response can be read out optically. The microwave signal is applied to the sample via a straight copper wire of $\SI{20}{\mu m}$ diameter. For fluorescence collection, we employ a high-NA objective (0.95 NA, $100\times$) with a diffraction limited resolution of $\SI{340}{nm}$. Figure~\ref{fig-one}(a) shows a confocal image of the diamond (110) side facet where the NVs are shallow implanted. In the region shown in Figure~\ref{fig-one}(a), there are no waveguides, letting us characterize the NV's properties without strain effects induced by the waveguides \cite{sotillo_polarized_2018}.

A photon shot-noise-limited sensitivity for an ensemble of NVs using CW-ODMR is given by~\cite{dreau_avoiding_2011}:

\begin{equation}
  \eta_{\mathrm{cw}} = \frac{4}{3\sqrt{3}}\frac{\Delta \nu}{\gamma C \sqrt{N}}   
  \label{eq:sensitivity}
\end{equation}

with $\frac{4}{3\sqrt{3}}$ a prefactor for a Lorenzian fitting ODMR for the maximum slope, $\mathrm{\Delta \nu}$ the linewidth, $\gamma = 2\pi \times \SI{28}{GHz T^{-1}}$ the electron spin gyromagnetic ratio, C the contrast and $N$ the fluorescence count rate. The minimum achievable linewidth, $\Delta \nu \approx \frac{1}{\pi T_2^*}$, is limited by the coherence time, ${T_2^*}$, of the NV electron spin \cite{dreau_avoiding_2011}.

The CW-ODMR spectrum at zero external magnetic field shown in Figure~\ref{fig-one}(b) with a single Lorenzian profile with the resonance frequency of $\SI{2.87}{GHz}$, yields a sensitivity value of $\SI{32}{\mathrm{nT}/\sqrt{\mathrm{Hz}}}$ using Eq.~\eqref{eq:sensitivity}. By applying an external magnetic field along \(\langle 100 \rangle\) direction, we split the resonance frequency of ${m_s = \pm 1}$ states and see two resonance dips separated by $\SI{148}{MHz}$ due to the Zeeman effect (Figure~\ref{fig-one}(b)). We then perform Rabi driving between ${m_s=0}$ and ${m_s=+1}$ states at resonance frequency $\SI{2.95}{GHz}$ shown in Figure~\ref{fig-one}(c). The Rabi oscillation has a $\SI{4.5}{MHz}$ frequency with contrast and lifetime of $\SI{11.9 \pm 0.5}{\%}$ and $\SI{570 \pm 50}{ns}$, respectively. The Ramsey free induction decay with the Rabi period of \SI{222}{ns} gives a $T_2^*$ = \SI{60.4 \pm 4.5}{ns} (Figure~\ref{fig-one}(d)). We attribute the fast dephasing time to the high concentration of implanted NVs ($75 \pm 25 \times 10^{9}$  $\text{NV}/\mathrm{cm^2}$) and dipole interaction among the centers themselves \cite{bauch_decoherence_2020} and the strain effect due to the proximity of NVs to the surface \cite{chakravarthi_impact_2021}. Ramsey measurements on NV- centers distant from the waveguide resulted in comparable coherence times, indicating that the waveguide-induced strain does not have a significant contribution to the coherence time, in agreement with Ref.\cite{guo_laser-written_2024}. Another decoherence mechanism is the inhomogeneity of the MW field, which can be improved by using an omega microwave structure~\cite{rezinkin_uniform_2024}. 

\begin{table}[ht]
\centering
\caption{\justifying {B-field sensitivity in ({${\mathrm{nT}/\sqrt{\mathrm{Hz}}}$}) for the marked waveguide (WG) set in Figure~\ref{fig-two}}.}
\label{your-table-i}

\begin{tabular}{c||l|l|l|l|l}

      row &WG1&WG2&WG3&WG4&WG5       \\ \hline \hline                        
$\mathrm{ I}$  & $338 \pm 17$ & $370 \pm 20$ & $130 \pm 5$ & $141 \pm 5$ & $145 \pm 5$  \\ \hline
$\mathrm{II}$ & $278\pm20$ & $208 \pm 15$ & $260 \pm12$ & $180\pm9$  & $213\pm11$  \\ \hline
$\mathrm{III}$& $79\pm4$  & $90\pm4$  & $129\pm8$ & $182\pm10$ & $179 \pm 8$ 

\end{tabular}

\end{table}
Next, we investigate the waveguide sensing properties by coupling the excitation and detection of NV fluorescence into the waveguides and performing CW-ODMR. The primary differences between confocal and waveguide-assisted sensing lie in the mode volume, the number of NV centers, and the excitation power. Waveguide-assisted ODMR excites NV centers within a $1/e^2$ Gaussian mode area of $\SI{456}{\mu m^2}$, resulting in a number of NV centers contributing to the ODMR signal that is about three orders of magnitude larger than those in the confocal configuration, which has an excitation area of $\SI{0.366}{\mu m^2}$. Consequently, higher excitation power is required in waveguide-assisted sensing to excite the larger mode area compared to confocal microscopy efficiently.

Figure~\ref{fig-two}(a) presents a microscope image of an array of studied waveguides on the (110) side facet of the diamond, where the NV centers are implanted. A 532 nm laser is coupled from the opposite side facet of the diamond using a low NA (0.25 NA, $10\times$) objective to excite the waveguide mode. The light travels through the waveguide to reach the facet with the implanted NV centers. The fluorescence signal then travels back through the waveguide and is collected using the same objective as for the excitation. Figure~\ref{fig-two}(b) displays the Gaussian mode of one of the studied waveguides, which has a $1/e^2$ area of $\SI{456}{\mu m^2}$, corresponding to the NV excitation area.
To ensure proper excitation and detection through the waveguide, we rotate a half-wave plate (HWP) in the excitation path and examine the NV's fluorescence spectrum [see Figure~\ref{fig-two}(c)]. It can be seen that when the laser polarization is perpendicular to the $y$-axis (parallel to waveguide sidewalls), the signal is minimally guided through the waveguide. This behavior reflects two distinct effects. First, the guidance of the laser depends on the alignment of its polarization with the fixed polarization of the waveguide. When the excitation laser's polarization is parallel to the waveguide’s polarization, we observe a transmission efficiency of $12-13\%$, along with a clear Gaussian mode profile. Conversely, when the laser polarization is perpendicular to the waveguide’s polarization, the transmitted power drops to $2-3\%$, and the Gaussian mode profile disappears. As a result, NV centers are excited with significantly reduced power in this perpendicular configuration, leading to a weaker photoluminescence (PL) signal.
Second, the fixed polarization of the waveguide not only defines the polarization of the light field with respect to the NV centers’ dipole orientations, but also affects the guidance of the NV dipole emission through the waveguide. Due to the varying crystallographic orientations of the NV centers, different dipoles exhibit different coupling efficiencies to the waveguide. This inherent property of the ensemble, combined with the fixed waveguide polarization, prevents the light field polarization from being optimized for all four NV dipole orientations simultaneously.
 Upon confirming proper excitation via the waveguides, we perform CW-ODMR on various waveguides, with one example shown in Figure~\ref{fig-two}(d). The sensitivity values for the set of waveguides marked in Figure~\ref{fig-two}(a) are presented in Table I, ranging from 79 to $\SI{370}{\mathrm{nT}/\sqrt{\mathrm{Hz}}}$. The average sensitivity across all the waveguides is $\SI{195 \pm 3}{\mathrm{nT}/\sqrt{\mathrm{Hz}}}$, with a standard deviation of $\SI{85}{\mathrm{nT}/\sqrt{\mathrm{Hz}}}$. This variation is attributed to differences in waveguide losses and the varying coupling efficiency of the NV dipoles to the waveguide mode.

\subsection{DC current vector sensing}\label{sec-vector-sensing}

\begin{figure}
\centering
\hspace{0cm}{\includegraphics[scale=0.61]{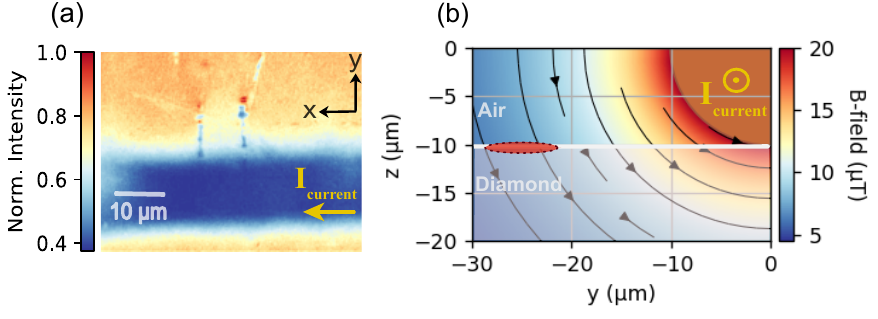}} 
\caption{\justifying \textbf{Magnetic field vector on one studied waveguide by applying a DC current.} (a) Confocal image of one of the studied waveguides in top view. The strip line area is the copper wire (as in Figure~\ref{fig-two}a) for applying the MW field and the DC-current. The $x$-axis is in the direction of the current flow. (b) A side-view magnetic field vector simulation (using Eq.~\eqref{biosavart}) from a DC carried by a $\SI{20}{\mu m}$ diameter wire. The y and $z$-axis shows the distance from the center of the wire carrying current along the $x$-axis. The red area with a dashed line is where the waveguide mode is, and $z = \SI{-10}{\mathrm{\mu m}}$ is the (110) surface of the diamond with the shallow implanted NVs a few nanometers below the surface.}\label{fig-dc-3}
\end{figure}

To benchmark the sensitivity performance of the waveguide array, a DC current is sent through a copper wire to the sample. Figure~\ref{fig-dc-3}(a) shows one of the studied waveguides in the confocal scan on the output of the waveguide where the NV centers are implanted near the copper wire. According to Biot-Savart law, the magnetic field generated by a constant electric current $I$ is:

\begin{equation}
    \vec{B} = \frac{\mu_0}{4\pi} \int \frac{\vec{I} \times \hat{r}}{r^2} \, dl
    \label{biosavart}
\end{equation}

with vacuum magnetic permeability of
    $\mu_0=4\pi\times10^{-7} \, \text{T/A}$

The center of the waveguide mode is at a distance ($r$) of $\SI{27}{\mathrm{\mu m}}\pm\SI{5.5}{\mu m}$ from the center of the wire. The large uncertainty is due to the waveguide mode area. Assuming a current of $\mathrm{I=\SI{1}{mA}}$, the magnetic field from Eq.~\eqref{biosavart} gives a value of $\sim \SI{7.43}{\mathrm{\mu T}}\pm \SI{2.3}{\mathrm{\mu T}} $. In Figure~\ref{fig-dc-3}(b), we map the magnetic field vector near the wire to determine the expected value using Eq.~\eqref{biosavart}.

\begin{figure*}
\centering
\includegraphics[scale=0.6]{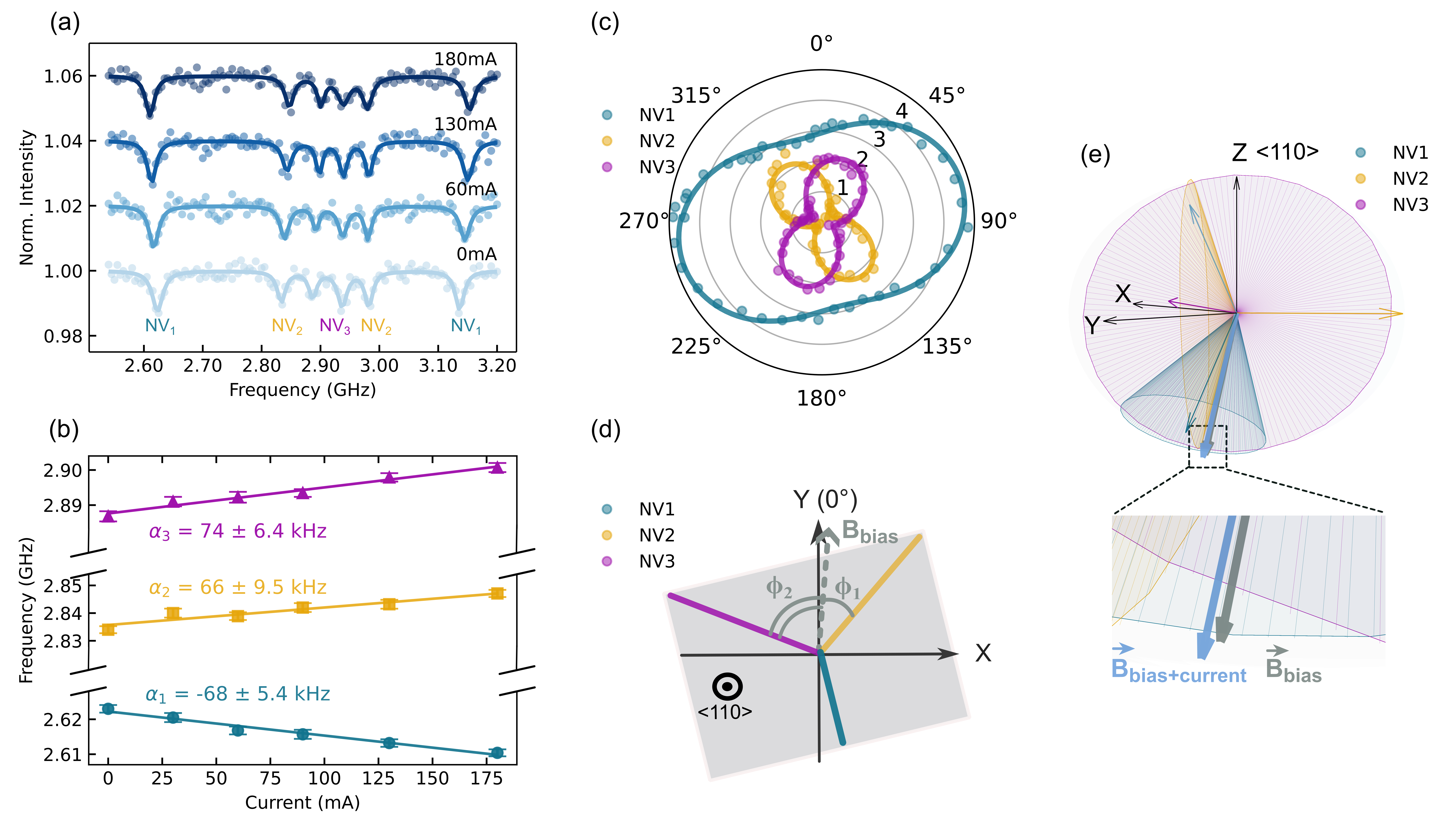} 
\caption{\justifying \textbf{B-field vector reconstruction of the DC current using three different NV crystallographic orientations.} (a) ODMR with an external magnetic field with different magnetic field projections on three NV orientations for four currents from 0-180 mA. Here the HWP in the excitation path is along Y-axis as shown in (d). (b) The resonance frequency shift of the three NVs with applied current for $m_s= 0$ to $m_s= -1$ transition. The data points are shown with blue circles for $\mathrm{NV_1}$, yellow squares for $\mathrm{NV_2}$, and purple triangles for $\mathrm{NV_3}$, and the slope ($\alpha$) of their linear fits in solid lines give the frequency shift per 1mA current. (c) polar plot of 3 NVs electric dipoles derived using a HWP in the excitation path within the confocal setup and measuring the contrast of ODMR signal; The $\theta_{light field} =0^{\circ}\pm5^{\circ}$ is parallel to the waveguide mode polarization axis. (d) Using the polar plot in (c), we show the schematic view of NV-axis orientations viewed along $z$-axis to (110) surface. The $y$-axis is along the waveguide polarization axis in the lab frame. The projection of the $B_{\mathrm{bias}}$ on (110) plane is shown with a dashed arrow with the azimuthal angle ${\phi_1= 33 \pm 3}^{\circ}$ and $\phi_2= {77 \pm 3}^{\circ}$ derived from Eq.~\eqref{azimuthal}. (e) Polar cones around three NV axes at $\mathrm{I = \SI{0}{mA}}$ showing possible orientations of the B-field vector from the bias field. The x-y plane is the (110) surface where two of the NV orientations (labeled as $\mathrm{NV_{2,3}}$) lie on it, and two other NV axes are $\pm 54.73^{\circ}$ out of the x-y plane.  The intersection of the three cones gives the B-field vector orientation. The zoom-in area shows the triangle area of the crossed cones with the B-field vectors of $B_{\mathrm{bias}}$ and $B_{\mathrm{bias+current}}$.} \label{fig-vector}
\end{figure*}

To investigate the DC current value and determine the vector field from the current, we measure CW-ODMR with an external bias magnetic field using a permanent magnet such that its projection on each NV orientation varies. This variation allows us to differentiate the contributions from the distinct NV crystallographic orientations and provides a mapping of the vector magnetic field generated by the DC current. Figure~\ref{fig-vector}(a) shows the waveguide-assisted ODMR spectra for four different currents, each displaying six distinct resonance dips. These dips correspond to the three different orientations of nitrogen-vacancy (NV) centers, labeled as $\mathrm{NV_{1,2,3}}$. While four NV families are theoretically expected, the absence of a pronounced dip for the fourth family is likely attributed to a combination of poor coupling to the RF signal and inhomogeneous broadening induced by local strain effects. These strain effects, caused by the waveguides, vary with NV orientation and can lead to broadening of the ODMR signal. Further details on the strain effects in waveguides can be found in Ref.~\cite{alam_determining_2024}

To estimate the B-field projection per 1 mA on each NV axis, the resonance shift value of ${m_s=-1}$ state for $\mathrm{NV_{1,2,3}}$ are depicted in Figure~\ref{fig-vector}(b). From the linear fits, the slope values $\alpha_{1,2,3}$ are extracted, yielding $\SI{- 68\pm 5.4}{\mathrm{kHz/mA}}$, $\SI{66\pm 9.5}{\mathrm{kHz/mA}}$ and $\SI{74\pm 6.4}{\mathrm{kHz/mA}}$, respectively. Using the formula $\delta \nu = \gamma B_z$ with $B_z$the projection of the B-field on NV-axis, these slopes corresponds to magnetic field values of \SI{2.43 \pm 0.19}{\mathrm{\mu T/mA}}, \SI{2.36 \pm 0.34}{\mathrm{\mu T/mA}} and \SI{2.64 \pm 0.23}{\mathrm{\mu T/mA}}.

Now, to fully characterize the B-field magnitude and orientation, we employ the following formula from  \cite{balasubramanian_nanoscale_2008}:

\begin{equation}
\left( g\mu_\text{B}B \right)^2 = \frac{1}{3} \left(\nu_1^2 + \nu_2^2 - \nu_1\nu_2 - D^2\right) - E^2,
\label{B-magnitude}
\end{equation} 

with $D = \SI{2.872}{GHz}$ and $E = \SI{8.15}{MHz}$ derived from the ODMR signal at zero B-field. 
To reconstruct the full B-field vector, we must also determine its orientation, including the polar angle ($\theta$) and the azimuthal angle ($\phi$). The polar angle $\theta$ is given by \cite{balasubramanian_nanoscale_2008}:

\begin{equation}
\resizebox{\columnwidth}{!}{%
$ \Delta = \frac{7D^3 + 2(v_1 + v_2)\left(2(v_1^2 + v_2^2) - 5v_1v_2 - 9E^2\right) - 3D\left(v_1^2 + v_2^2 - v_1v_2 + 9E^2\right)}{9\left(v_1^2 + v_2^2 - v_1v_2 - D^2\right) - 3E^2} $
}
\label{B-polar}
\end{equation}

with $\Delta \approx D\cos(2\theta)$.

Table \ref{your-table-ii} (\ref{your-table-iii}) shows the magnitude of the B-field and its polar angles at $\mathrm{I = \SI{0}{mA}}$ ($\mathrm{I = \SI{30}{mA}}$) with $\left|B\right|cos\theta$ being the projection of the B-field on each NV-axis. From the difference of $B (\mathrm{I=30}) - B (\mathrm{I=0})$, the B-field generated by the DC current of 30 mA can be determined as \SI{0.21 \pm 0.16}{mT}. This value closely aligns with the simulated B-field value of \SI{0.22 \pm 0.07}{mT}, as illustrated in Figure~\ref{fig-dc-3}, validating the experimental results.

\begin{table}[ht]
\centering
\caption{\justifying {B-field magnitude and polar angle at I = 0 mA.}}
\label{your-table-ii}

\begin{tabular}{cccccc}

    &$B$ (mT)&$\theta ^{\circ}$ & ${\left|B\right|\mathrm{cos\theta}~(mT)}$ \\ \hline\hline
$\mathrm{NV_1}$  & $10.15 \pm 0.09$ & $25.4\pm 1.1$ & $9.18 \pm 0.12$  \\ \hline
$\mathrm{NV_2}$   & $9.75 \pm0.11$ & $74.36\pm0.27$ & $2.62 \pm 0.05$  \\ \hline
$\mathrm{NV_3}$ & $9.95 \pm 0.12$ & $85.61 \pm0.26$ & $0.726 \pm 0.046$\\ 

\end{tabular}

\end{table}

\begin{table}[ht]
\centering
\caption{\justifying {B-field magnitude and polar angle at I = 30 mA.}}
\label{your-table-iii}
\begin{tabular}{cccccc}

     &$B$ (mT)&$\theta ^{\circ}$ & ${\left|B\right|\mathrm{cos\theta}~(mT)}$ \\ \hline\hline
$\mathrm{NV_1}$  & $10.18\pm0.1$& $24.6\pm1.2$ & $9.24 \pm 0.13$  \\ \hline
$\mathrm{NV_2}$  & $10.03\pm0.11$ & $75.69\pm0.32$ & $2.48 \pm 0.06$ \\ \hline
$\mathrm{NV_3}$  & $10.26\pm0.13$ & $86.15\pm0.26$  & $0.677 \pm 0.047$ \\ 

\end{tabular}

\end{table}

Next, we calculate the azimuthal angle ($\phi$) by knowing the orientation of each NV-axis with respect to the (110) facet. Two possible NV orientations are parallel to the (110) plane, and the other two are tilted $\pm 54.73$ degrees out of the plane.\\ 
To identify which resonance dips in Figure~\ref{fig-vector}(a) correspond to the in-plane NV axis, we perform ODMR in a confocal setup with a half-wave plate (HWP) in the excitation path and calculate the contrast of each resonance dip while rotating the polarization of the excitation light field~\cite{alegre_polarization-selective_2007}. Figure~\ref{fig-vector}(c) shows the polar plot of three NV orientations labeled as $\mathrm{NV_{1,2,3}}$.

The polar plots for $\mathrm{NV_{2}}$ and $\mathrm{NV_{3}}$ indicate a single electric dipole pattern, while $\mathrm{NV_{1}}$ displays a pattern consistent with two electric dipoles. Each NV center has a pair of orthogonal dipoles that are perpendicular to the spin axis (NV-axis). Considering the (110) crystal facet and the two in-plane NV orientations, each in-plane NV has a pair of dipoles, with one dipole from each pair not coupling to the (110) plane.
Therefore, based on the dipole patterns observed in Figure~\ref{fig-vector}(c), $\mathrm{NV_{2}}$ and $\mathrm{NV_{3}}$ are identified as the in-plane NV axes. Conversely, the $\mathrm{NV_{1}}$ pattern suggests it corresponds to one of the out-of-plane NV orientations. The $\mathrm{NV_{1}}$ orientation is explained later in the text.

Using the B-field projections from Tables \ref{your-table-ii} and \ref{your-table-iii}, we can calculate the azimuthal angle $\phi$ using the following equation:

\begin{equation}
\frac{\cos(\phi_1)}{\cos(\phi_2)} = \frac{B \mathrm{\cos\theta_2}}{B \mathrm{\cos\theta_3}}
\label{azimuthal}
\end{equation}

with $\phi_2=109.47^{\circ} - \phi_1$ and $\theta_{2,3}$ are the polar angle of $\mathrm{NV_{2,3}}$ respectively.

We derive the following azimuthal angles: ${\phi_1= 33 \pm 3}^{\circ}$ and $\phi_2= {77 \pm 3}^{\circ}$. 

Another significant advantage of determining the relative dipole orientations by rotating the HWP is understanding the electric dipoles' orientation relative to the waveguide mode's polarization axis. As discussed in Figures~\ref{fig-two}(b) and \ref{fig-two}(c), the waveguides guide light with a polarization parallel to the waveguide sidewalls ($y$-axis). We used the same HWP orientation angle in the confocal setup for the polar pattern measurements as we did for the waveguide-assisted ODMR in Figure~\ref{fig-two}(e). In Figure~\ref{fig-vector}(c), the polarization axis of the light field corresponds to ${0\pm 5}^\circ$, which is parallel to the $y$-axis in Figure~\ref{fig-vector}(d). This approach allows us to map the B-field comprehensively in the lab frame with respect to the waveguide mode polarization axis.  

In Figure~\ref{fig-vector}(e), we plot the polar cones around the three NV axes to visualize the B-field vector. These cones represent the possible orientations of the B-field based on the polar angles $\theta$ from Table~\ref{your-table-ii}. The intersection of these three cones determines the B-field orientation in NV crystal frame. Geometrically the cones do not intersect if $\theta_i + \theta_j < 109.47^\circ$. Thus, if $\theta_i > 54.735^\circ$, we perform a point mirroring with respect to the origin and define a new angle as $\theta'_i = 180^\circ - \theta_i$ \cite{weggler_determination_2020}. In Figure~\ref{fig-vector}(e), the polar cones for $\mathrm{NV_{2,3}}$ are point mirrored relative to the origin. 

To determine whether $\mathrm{NV_{1}}$ is oriented at $54.73^{\circ}$ outward from the (110) surface or inward towards the (110) surface, we refer to the simulation in Figure~\ref{fig-dc-3}. Given that the B-field from the current is directed inward towards the (110) surface, we conclude that $\mathrm{NV_{1}}$ is oriented inward towards the (110) facet. Consequently, the intersection of the three cones will be inward towards the (110) surface. Ideally, the intersection of the three cones would form a single point, indicating the precise location of the B-field vector. However, due to experimental uncertainties such as power broadening of the ODMR resonance dips, lattice strain, and temperature fluctuation during the measurement process, the cones intersect to form a triangle~\cite{chen_calibration-free_2020}. We use the center of this triangle to determine and plot the B-field vector.



\section{Experimental Setup} \label{sec-setup}
The femtosecond laser employed for waveguide writing in diamond is a Yb: KGW fiber laser (Bluecut, Menlosystems) with a 300-fs pulse duration and a 515-nm wavelength (frequency-doubled using an LBO crystal). The laser is focused using a 1.25-NA oil immersion lens (RMS100×-O ×100 Olympus Plan Achromat Oil Immersion Objective, 100× oil immersion, Olympus). Polished synthetic single-crystal diamond samples ($\SI{2}{mm} \times \SI{2}{mm} \times \SI{0.5}{mm}$, type IIa, electronic grade with nitrogen impurities $<$ 5 ppb, Element 6) are utilized. The sample is mounted on a computer-controlled, three-axis motion stage (ANT130 series, Aerotech) to translate the sample relative to the laser, forming the desired photonic structures. The polarization of the incident laser is oriented perpendicular to the scan direction. The NV centers fabricated by shallow implantation of $[^{15}\text{N}^+]$ into (110) facets. Using a homebuilt low-energy ion implanter equipped with a Wien mass filter as well as an Einzel lens for beam focusing, nitrogen ions of energy 2.5 or 5 keV implanted at doses of $2,4,5 \times 10^{12}$  $^{15}\text{N}^+/\mathrm{cm^2}$. For implantation energies of 2.5 keV and 5 keV, the NV creation yields is about \(1.5 \pm 0.5\%\) and \(3.5 \pm 1.0\%\), respectively, resulting in NV densities of \((75 \pm 25) \times 10^9 \, \text{NV/cm}^2\) and \((140 \pm 40) \times 10^9 \, \text{NV/cm}^2\). Subsequently, the substrate is annealed in ultra-high vacuum (UHV) for 3 hours at $1000^\circ\mathrm{C}$. More info on the implantation procedure can be found in \cite{lang_long_2020}.

The CW-ODMR is measured using a microwave source (Rohde \& Schwarz SMIQ 03), connected to an amplifier (mini-circuits ZHL-42+), and a final output power of \SI{28}{dBm}. The DC current is created by a Source Measure Unit (SMU) and connected to a BiasTee (mini-circuits ZX85-12G-S+) to be combined with the microwave field and carried out to the sample with a copper wire of $\SI{20}{\mu m}$ diameter. A CW green laser at \SI{532}{nm} wavelength was used for optical pumping.


\section{Summary} \label{sec-summary}

In this study, we explore integrating shallow-implanted NV centers in diamond with laser-written waveguides, aiming to enhance the performance and scalability of quantum sensing devices. Our approach leverages the control over light propagation paths provided by femtosecond laser-written waveguides to facilitate excitation and fluorescence detection from an ensemble of NV centers located 8-12 nm below the diamond surface, enable sensing of bio-samples positioned in close proximity to the diamond within an area of about $\SI{456}{\mu m ^2}$.
We first characterized the NV centers using a home-built confocal microscope, evaluating their magnetic field sensitivity and coherence properties through CW-ODMR and pulsed measurements. We further investigated the performance of waveguide-assisted ODMR. Exciting NV centers through the waveguides resulted in larger mode volumes and higher excitation power requirements, resulting in sensitivity values ranging from 79 to $\SI{370}{\mathrm{nT}/\sqrt{\mathrm{Hz}}}$ depending on waveguide losses and NV coupling efficiency. However, sensitivities in the picotesla to the femtotesla range have been demonstrated in bulk diamonds using techniques like confocal microscopy and widefield imaging~\cite{PhysRevApplied.22.044069, PhysRevX.5.041001}. These results highlight potential pathways for further advancements in NV-center creation and ODMR detection strategies. For instance, lock-in detection or pulsed-microwave sequences could significantly enhance sensitivity in future implementations~\cite{RevModPhys.92.015004}.
We applied an external DC current to validate the sensing capabilities and observed the resonance shift in the ODMR spectrum. The magnetic field generated by the current was mapped using three different NV crystallographic orientations, allowing for a comprehensive reconstruction of the B-field vector in both the crystal and lab frames. This mapping utilized the distinct polarization axes of the waveguides and NV centers, enabling precise B-field vector determination. The demonstrated DC magnetic field detection and vector reconstruction serve as a proof of principle for future applications, such as mapping complex magnetic fields like those generated by neuronal action potentials~\cite{barry_optical_2016}.

Our findings demonstrate the potential of integrating shallow-implanted NV centers with laser-written waveguides for advanced quantum sensing applications. In particular, we focused on reconstructing the magnetic field vector using a single waveguide sensing task. This reconstruction is performed with respect to the waveguide sidewall aligned with the polarization axis of the waveguides, which is distinct from the intrinsic NV orientations, a property that can be accessed visually from the sample. This approach paves the way for developing scalable, efficient, and easy-to-use quantum sensing devices. This work underlines the promising synergy between NV center technology and femtosecond laser fabrication, contributing to the ongoing advancement in quantum information processing and sensing technologies.

\begin{acknowledgments}
S.S., G.C., R.R., A.J.B., J.P.H., S.M.E. and A.K. acknowledges support of the Marie Curie ITN project LasIonDef (GA n.956387). S.M.E. is thankful for the support from the projects QuantDia (FISR2019-05178) and PNRR PE0000023 NQSTI funded by MUR (Ministero dell’Università e della Ricerca). A.J.B. and J.P.H. acknowledge the financial support provided by EPSRC via Grant No. EP/T017813/1 and EP/03982X/1.
Most measurements were performed on the basis of the Qudi software suite.\cite{binder_qudi_2017}
\end{acknowledgments}

\subsection*{Author Contributions}
\textbf{Sajedeh Shahbazi}: Conceptualization (equal); Data curation (equal); Investigation (equal); Methodology (equal); Visualization (equal); Writing – original draft (equal); Writing – review $\And$ editing (equal). \textbf{Giulio Coccia}: Methodology (equal); Writing – review $\And$ editing (equal). \textbf{Argyro N. Giakoumaki}: Methodology (equal). \textbf{Johannes Lang}: Methodology (equal); \textbf{Vibhav Bharadwaj}: Methodology (equal). \textbf{Fedor Jelezko}: Writing – review $\And$ editing (equal). \textbf{Roberta Ramponi}: Funding acquisition (equal); Writing – review $\And$ editing (equal).  \textbf{Anthony J. Bennett}: Funding acquisition (equal); Writing – review $\And$ editing (equal). \textbf{John P. Hadden}: Funding acquisition (equal); Writing – review $\And$ editing (equal). \textbf{Shane M. Eaton}: Funding acquisition (equal); Supervision (equal); Writing – review $\And$ editing (equal). \textbf{Alexander Kubanek}: Conceptualization (equal); Funding acquisition (equal); Supervision (equal); Writing – review $\And$ editing (equal).

\bibliography{aipsamp}  

\end{document}